\documentclass[aps,prl,twocolumn,showpacs,superscriptaddress]{revtex4-1}
\usepackage{graphicx,amsmath,amssymb,amsfonts}
\usepackage[latin1]{inputenc}
\usepackage{array}
\usepackage{multirow}
\usepackage{float}
\usepackage{color}
\usepackage{braket}
\usepackage[normalem]{ulem}
\begin{document}
\title{Antiferromagnetic structure and magnetic properties of Dy$_{2}$O$_{2}$Te: an isostructural analog of the rare-earth superconductors {\it R}$_{2}$O$_{2}$Bi }

\author{Juanjuan Liu}
\author{Jiale Huang}
\affiliation{Laboratory for Neutron Scattering and Beijing Key Laboratory of Optoelectronic Functional Materials and MicroNano	Devices, Department of Physics, Renmin University of China,	Beijing 100872, China}

\author{Jieming Sheng}
\affiliation{Department of Physics, Southern University of Science and Technology, Shenzhen 518055, China}
\affiliation{Institute of High Energy Physics, Chinese Academy of Sciences (CAS), Beijing 100049, China}
\affiliation{Spallation Neutron Source Science Center (SNSSC), Dongguan 523803, China}

\author{Jinchen Wang}
\author{ Feihao Pan}
\affiliation{Laboratory for Neutron Scattering and Beijing Key Laboratory of Optoelectronic Functional Materials and MicroNano	Devices, Department of Physics, Renmin University of China,	Beijing 100872, China}

\author{Hongxia Zhang}
\author{Daye Xu}
\affiliation{Laboratory for Neutron Scattering and Beijing Key Laboratory of Optoelectronic Functional Materials and MicroNano	Devices, Department of Physics, Renmin University of China,	Beijing 100872, China}

\author{Jianfei Qin}
\author{Lijie Hao}
\affiliation{ China Institute of Atomic Energy, PO Box-275-30, Beijing 102413, China}

\author{Yuanhua Xia}
\author{Hao Li}
\affiliation{ Key Laboratory of Neutron Physics and Institute of
	Nuclear Physics and Chemistry, China Academy of Engineering
	Physics, Mianyang 621999, China}

\author{Xin Tong}
\affiliation{Institute of High Energy Physics, Chinese Academy of Sciences (CAS), Beijing 100049, China}
\affiliation{Spallation Neutron Source Science Center (SNSSC), Dongguan 523803, China}

\author{Liusuo Wu}
\affiliation{Department of Physics, Southern University of Science and Technology,
	Shenzhen 518055, China}

\author{Peng Cheng}
\email[Corresponding author: ]{pcheng@ruc.edu.cn}
\affiliation{Laboratory for Neutron Scattering and Beijing Key Laboratory of Optoelectronic Functional Materials and MicroNano	Devices, Department of Physics, Renmin University of China,	Beijing 100872, China}

\author{Wei Bao}
\email[Corresponding author: ]{weibao@cityu.edu.hk}
\affiliation{Department of Physics, City University of Hong Kong,
	Kowloon, Hong Kong SAR}
\affiliation{Center for Neutron Scattering, City University of Hong Kong,
	Kowloon, Hong Kong SAR}

\begin{abstract}
The rare-earth compounds {\it R}$_{2}$O$_{2}$Bi ({\it R} = Tb, Dy, Er, Lu, Y) are newly discovered
superconductors in the vicinity of a rare-earth magnetic long-range order. 
In this work,
we determined the magnetic order of the parent compound Dy$_{2}$O$_{2}$Te by neutron scattering as the $A$-type antiferromagnetic structure below the N\'{e}el temperature $T_N=9.7~$K.  
The large staggered magnetic moment 9.4(1)$ \mu_{B} $ per Dy at $T = 3.5~$K lies in the basal $ab$-plane. 
In a magnetic field, anomalous magnetic properties including the bifurcation between zero-field and field cooling magnetization,
a butterfly-shaped magnetic hysteresis and slow magnetic relaxation emerge, which are related to the field-induced metamagnetic transitions in Dy$_{2}$O$_{2}$Te. 
Our experimental findings would stimulate further research on the relation between antiferromagnetism and superconductivity in these rare-earth compounds.
\end{abstract}

\maketitle


\section{Introduction}
The layered compounds of the ThCr$_{2}$Si$_{2}$ structure have shown great variety of novel superconducting and magnetic properties. 
The famous examples include the first heavy-fermion superconductor CeCu$_{2}$Si$_{2}$ discovered in 1979 \cite{PhysRevLett.43.1892}  and subsequent isostructural heavy-fermion superconductors \cite{Jaccard_1992,roma_1996, Gros_1996, u122_palstra}.
More recently in 2008 Fe-based superconductors BaFe$_{2}$As$_{2}$, CaFe$_{2}$As$_{2}$ and SrFe$_{2}$As$_{2}$ of the ThCr$_{2}$Si$_{2}$ structure were discovered \cite{Ba122,Gen_Fu_2008,Sr122,Wu_2008,Ca122_pressure}. 
Different from conventional superconductivity which are caused by the electron-phonon interaction as explained by the BCS theory, magnetism is generally believed to play a crucial role in the unconventional superconductivity of the heavy fermion \cite{THOMPSON2003446,Stockert2011} and Fe-based \cite{QHuang_2008,PhysRevB.78.140504,PhysRevB.78.100506,Chen_2009,PhysRevLett.103.087001,PhysRevLett.101.057003,Dai2012,Hirschfeld_2011, Bao_2013,chubukov2012,rev2009hosono}
 superconductors.

{\it R}$_{2}$O$_{2}${\it X} ({\it R} = rare earth, {\it X} = Te, Bi, Sb) represents a large family of materials with the ThCr$_{2}$Si$_{2}$ crystal structure \cite{Te221, Bi221, Sb221}, which is composed of alternating [{\it R}$_{2}$O$_{2}$]$^{2+}$ and {\it X}$^{2-}$ stacking layers with opposite charges, in analog to the [Cr$_{2}$Si$_{2}$]$^{4-}$ and Th$^{4+}$ layers. 
The crystal structure is also referred to as the anti-ThCr$_{2}$Si$_{2}$-type for the switched signs of charges in the two layers. 
This kind of materials demonstrate numerous intriguing physical properties such as metal-insulator transition  \cite{Bi221_MIT,WANG2013,La221_O_2020}, charge density wave \cite{Bi221_CDW,CDW_Sb_2015} and Kondo effect \cite{ZAXu_2020}. Most interestingly, {\it R}$_{2}$O$_{2}$Bi with heavy rare-earth elements ({\it R} = Tb, Dy, Er, Lu, Y) were recently discovered to be superconducting below $\sim$2~K via excess oxygen incorporation \cite{Bi221_SC1,Bi221_SC2,Bi221_SC3}. 
Remarkably, Tb$_{2}$O$_{2}$Bi and Er$_{2}$O$_{2}$Bi are found to become antiferromagnetic in a first-ordered transition when they become superconductor, exhibiting the coexistence of antiferromagnetism with superconductivity according to the measurements of resistivity and susceptibility \cite{Bi221_MIT,Bi221_SC2,Bi221_SC3,qiao2021coexistence}. 

There have existed numerous discussions on the superconducting properties and mechanism of {\it R}$_{2}$O$_{2}$Bi. 
For example, the superconductivity is considered as arising from the metallic Bi square nets and the unit cell tetragonality (\textit{c/a}) is suggest to be closely related to superconducting transition temperature T$_c$ \cite{Bi221_SC3}. 
The superconducting Bi square nets are also associated with the possibilities of exploring topological superconductivity \cite{Bi221_SC2,Topo2015}. Theoretical calculations revealed suppression of the charge density wave instability in {\it R}$_{2}$O$_{2}$Bi due to large spin-orbit coupling, which may be associated with the emergence of superconductivity \cite{Bi221_CDW}. 
Additionally, some members of {\it R}$_{2}$O$_{2}$Bi can be considered as an alternating stacking of antiferromagnetic [{\it R}$_{2}$O$_{2}$]$^{2+}$ layers and superconducting Bi layers. 
The long-range ordered rare-earth antiferromagnetism in the [{\it R}$_{2}$O$_{2}$]$^{2+}$ layer naturally generates scientific interests on its relations with the superconductivity in the Bi layer. 
A very recent study on Er$_{2}$O$_{2}$Bi suggested that there may be a competition between superconductivity and antiferromagnetic order \cite{qiao2021coexistence}. 
Therefore, this series of materials provide a new platform for investigating the interplay between antiferromagnetism and superconductivity.

Antiferromagnetic properties of {\it R}$_{2}$O$_{2}$${\it X}$ compounds, however, have not been studied in depth. 
Take Dy$_{2}$O$_{2}$Te for instance, the only knowledge about its physical properties so far comes from an early M\"{o}ssbauer study, which shows that Dy$_{2}$O$_{2}$Te is an antiferromagnet with T$_N\sim 10~$K \cite{CHAPPERT1977102}. 
However the antiferromagnetic structure and detailed magnetic properties are still unknown. 
Since the Te- and Bi-series have identical [{\it R}$_{2}$O$_{2}$]$^{2+}$ layers, but the higher T$_N$ makes the former compound easier to access the ordered phase, the determination of the magnetic structure and the investigation of magnetic properties of {\it R}$_{2}$O$_{2}$Te would provide fundamental information on these exciting family of rare-earth material {\it R}$_{2}$O$_{2}${\it X}. Besides, due to the large quantum number $J$, many Dy-based materials exhibit peculiar magnetic properties such as slow magnetic relaxation behavior. This behavior has been widely observed in Dy-based complex in the research of single-molecule magnet or compounds with diluted Dy ions, which arises from the magnetic quantum tunneling effect and weak interactions between molecules\cite{SMM1_2003,SMM2_2009,SMM3_2014,SMM4_2020,1998}. The occurrence of slow magnetic behavior in a correlated Dy-based compound is very rare. A recent example is Dy$_2$Ti$_2$O$_7$, its interesting slow magnetic relaxation is possibly due to the special spin-ice state and dipolar spin correlations\cite{Dy227_2001,Dy227_2006}. It would also be interesting to explore whether this peculiar magnetic property would exist in Dy$_{2}$O$_{2}${\it X} series.

In this paper, we report the investigations on Dy$_{2}$O$_{2}$Te. We found that Dy$_{2}$O$_{2}$Te orders antiferromagnetically below the N\'{e}el temperature T$_N=9.7~$K. 
A collinear $A$-type magnetic structure with moments lie in the basal $ab$-plane is determined directly through powder neutron diffraction measurements.
The magnetization measurements reveal that Dy$_{2}$O$_{2}$Te exhibits magnetic-field-induced slow magnetic relaxation. Possible mechanisms such as the phonon bottleneck effect will be discussed. 

\section{Experimental methods}
Polycrystalline samples of Dy$_{2}$O$_{2}$Te were synthesized by
solid-state reaction of Dy granules and Te and Dy$_{2}$O$_{3}$
powders in stoichiometric portions. These reagents were mixed, pressed into pellets and
heated in an evacuated quartz tube at $1073~$K for $24~h$. 
The products were then reground, pressed into pellets and annealed at
$1223~$K for $20~h$. 
X-ray diffraction refinement indicates that samples prepared in such a way, 
referred to as Sample A in the remaining text, contain about 6-8\% weight fraction of 
unreacted Dy$_2$O$_3$ as an impurity phase. 
By adding extra amount of Te ($\sim$5\%) to the synthesis reagents, it is found that 
phase-pure Dy$_{2}$O$_{2}$Te samples can be obtained and they are referred to as Sample B. 
The magnetization measurements reported in this paper used 
Sample B, while the neutron diffraction experiments were performed on Sample A. 

Powder X-ray diffraction (XRD) patterns were collected
from a Bruker D8 Advance X-ray Diffractometer using Cu K$_{\alpha}$ radiation. 
Magnetization measurements were carried out in Quantum Design MPMS3 and PPMS-14T. 
The powder neutron diffraction experiments were carried out on Xingzhi triple-axis
spectrometer \cite{XingZhi} at the China Advanced Research Reactor (CARR) and Xuanwu powder neutron diffraction
spectrometer at China Academy of Engineering Physics (CAEP).
Approximate $3.8~$g of Dy$_{2}$O$_{2}$Te powder sample sealed in a
cylindrical vanadium or aluminium container was loaded into a
closed cycle refrigerator that regulates the sample temperature from $3.5~$K to $300~$K. 
For neutron experiments on Xingzhi, a neutron
velocity selector was used upstream to cleanly remove higher order neutrons for the
incident neutron energy fixed at $15$~meV \cite{XingZhi}.

\begin{figure}[bt] \centering
	\includegraphics[width=8.5cm]{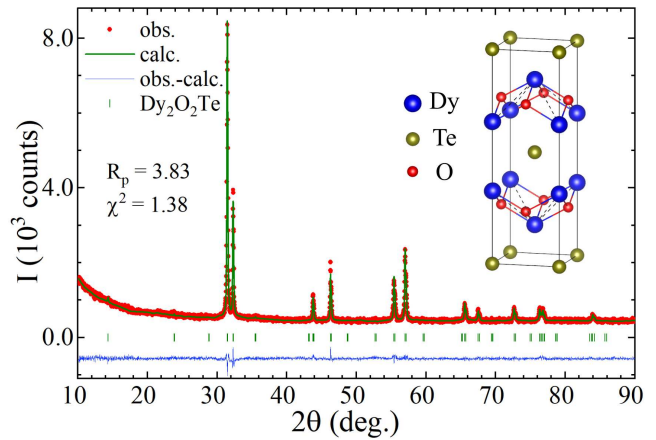}
	\caption {The x-ray powder diffraction pattern of Dy$_{2}$O$_{2}$Te
		measured on Sample B at room temperature. The inset shows the crystal structure
		of Dy$_{2}$O$_{2}$Te. } \label{XRD}
\end{figure}

The program FULLPROF Suite \cite{RODRIGUEZCARVAJAL199355} was used in the Rietveld refinement of neutron powder diffraction data. 
The aluminium peaks are accounted for by Le Bail fitting \cite{LEBAIL1988447}.
BasIreps program of FULLPROF Suite is used for representation
analysis to derive the possible magnetic structure modes. 
Because the dysprosium is highly neutron absorbing, the absorption
correction was applied to neutron powder diffraction data.

\section{Results and discussions}
\subsection{X-ray analysis and magnetic susceptibility}
From the powder x-ray diffraction pattern and its Rietveld
analysis shown in Fig. \ref{XRD}, the Sample B was confirmed to be phase-pure Dy$_{2}$O$_{2}$Te. 
Any possible impurities should be less than 1\%. Dy$_{2}$O$_{2}$Te
crystallizes in the tetragonal space group $I4 / mmm$. The
magnetic ions Dy occupy the Si sites of the ThCr$_{2}$Si$_{2}$
structure. The lattice parameters obtained from the refinement are
$a=b=3.925$ \text{\AA} and $c=12.413$ \text{\AA}. The nearest Dy-Dy distance is
$3.536 $ \text{\AA} which is indicated by the blue dashed line in Fig. \ref{XRD}.

\begin{figure}[bt] \centering
	\includegraphics[width=7.5cm]{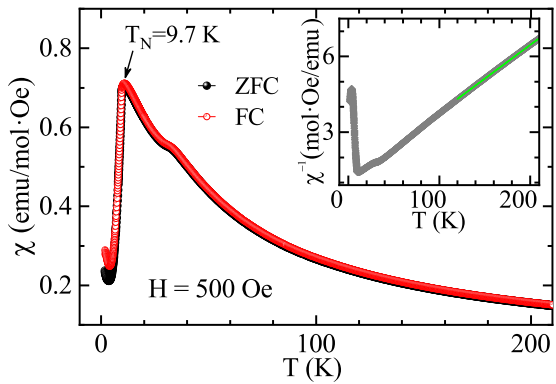}
	\caption {Zero-field-cooling (ZFC) and field-cooling (FC)
		magnetic susceptibility of Dy$_{2}$O$_{2}$Te at
		$H=500~$Oe. The inverse magnetic susceptibility $1/\chi$ versus temperature
		is plotted in the inset.} \label{MT}
\end{figure}

The temperature dependent magnetic susceptibility of
Dy$_{2}$O$_{2}$Te, measured at $H=500~$Oe, is presented in Fig. \ref{MT}. 
There is a cusp at $T_N = 9.7~$K in the magnetization
data indicating the occurrence of an antiferromagnetic transition. 
The susceptibility also exhibits an anomaly at about $30~$K. 
At temperatures above $40~$K, 
the linear relation between inverse magnetic susceptibility and temperature is apparent.
Fitting the high temperature data above 120~K to the Curie-Weiss law yields $\mu_{eff} = 8.6~\mu_{B}$ 
and $\theta_{CW} = -37.9~K$. 
The interpretation of the high temperature fitting results need to be careful in view of the crystal field 
scheme of the rare-earth ions in the material. 
We carried out the crystal field calculations based on the point-charge model 
using the software McPhase \cite{mcphase} for Dy$_{2}$O$_{2}$Te. 
The Dy$^{3+}$ ion has a ground state of the $16$-fold degenerate $J=15/2$ ($L=5$, $S=5/2$) multiplet of an effective magnetic moment 10.6~$\mu_{B}$.
The multiplet is further split into eight doublets by crystalline electrical field effect. The calculated first excited crystal field levels are at $1.95~$meV ($23~$K). Therefore in the high temperature region of Fig. \ref{MT}, the contributions from the excited states could be large 
and the Curie-Weiss fitting results for $\mu_{eff}$ and $\theta_{CW}$ does not represent the ground state magnetic properties at the base temperature. 
The susceptibility anomaly around $30~$K is likely caused by enhanced magnetic contributions from the thermal occupation of the $23~$K crystal field levels.
Indeed, evidence for the magnetic phase transition around $30~$K haven't been observed in our neutron diffraction experiments to be presented in the next part. 
The calculations base on point-charge model are rough evaluations. 
However, the preliminary crystal field 
scheme is sufficient for an understanding of the magnetic susceptibility data.

\begin{figure}[tb] \centering
	\includegraphics[width=8.5cm]{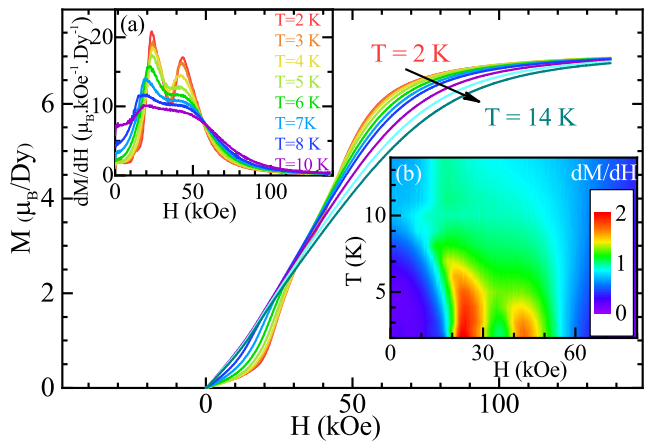}
	\caption {Magnetization isotherms measured at the specified temperatures. Initially, the data were measured after zero-field-cooling the sample to 2~K and with increasing field. After the measurement at 2~K, the field was cooling to zero at this temperature, then warming up directly to 3~K for next measurement, and so on. Insets: (a) The dM/dH curves at different temperatures. (b) Contour plot of dM/dH as a function of temperature and field.} \label{MH}
\end{figure}

Fig. \ref{MH} shows the isothermal magnetization measurements for
Dy$_{2}$O$_{2}$Te at various temperatures. With increasing the
applied magnetic field, two kinks are observed in M(H) at low temperatures. 
This behavior can be more clearly identified by two peaks at around $23~$kOe and $43~$kOe in 
the dM/dH curves as shown in the insets (a) of Fig. \ref{MH}. 
These two peaks gradually disappear with increasing temperature as illustrated 
by the contour plot of dM/dH in the insets (b) of Fig. \ref{MH}. 
Typically, these features are indications of metamagnetic transition\cite{PCheng_APL2020}, 
which will be further discussed in a following section. 
The magnetization approaches a saturated value $6.9~\mu_{B}/$Dy$^{3+}$ at 14~T and those measurement temperatures. This value is quite close to $2/3$ of the full saturation moment of Dy$^{3+}$ at the ground state ($\sim$10~$\mu_{B}$), which agrees well with the expectation of powder-averaged magnetization of an easy-plane magnet. In our McPhase calculations on Dy$_{2}$O$_{2}$Te, the crystal field $xyz$ axes are chosen to be along the crystal lattice $abc$ directions.
Then the calculated ground-state wave functions are:  
\[{E_{0 \pm }} =  \pm 0.9601\left| {{{ \pm 1} \mathord{\left/	{\vphantom {{ \pm 1} 2}} \right.\kern-\nulldelimiterspace} 2}} \right\rangle  \pm 0.2529\left| {{{ \mp 7} \mathord{\left/	{\vphantom {{ \mp 7} 2}} \right.	\kern-\nulldelimiterspace} 2}} \right\rangle  \pm 0.1198\left| {{{ \pm 9} \mathord{\left/	{\vphantom {{ \pm 9} 2}} \right.\kern-\nulldelimiterspace} 2}} \right\rangle .\]
This result suggests that the magnetic moment along the $c$-aixs would be quite small and it should lie in the $ab$-plane. The following determination of magnetic structure by neutron scattering also provides a consistent result.  

In addition, we measured the resistivity of Dy$_{2}$O$_{2}$Te at around room temperature which is about $6\times 10^3~\Omega \cdot$cm. The resistivity increases quickly with decreasing temperature and the data below $240~$K cannot be obtained due to the upper limit of PPMS measurement. Thus Dy$_{2}$O$_{2}$Te is a semiconductor, the same as previously reported for other members of {\it R}$_{2}$O$_{2}$Te (R = La, Sm, Gd) \cite{Te221_chi}. 
  
\begin{figure}[tbhp] \centering
	\includegraphics[width=8.5cm]{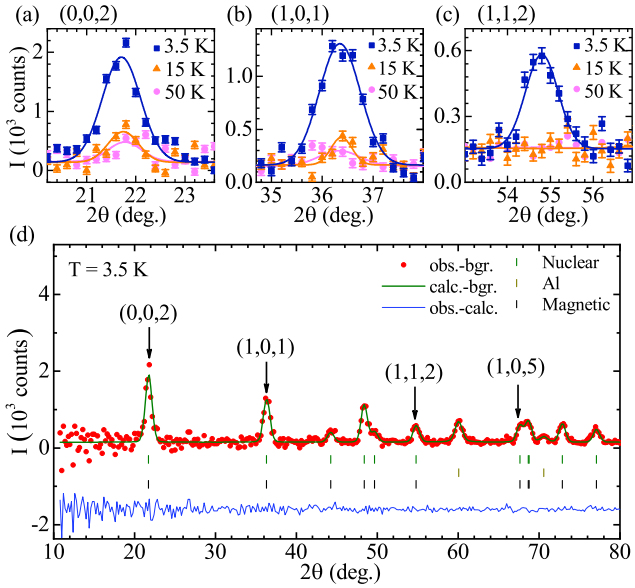}
	\caption {(a)$-$(c) Neutron diffraction peak profiles for (0,0,2), (1,0,1) and
		(1,1,2) at selected temperatures. (d) The powder neutron diffraction pattern of Dy$_{2}$O$_{2}$Te at 3.5 K and the Rietveld refinement fit.} \label{NPD}
\end{figure}

\subsection{Magnetic structure}
\begin{figure}[bt] \centering
	\includegraphics[width=8cm]{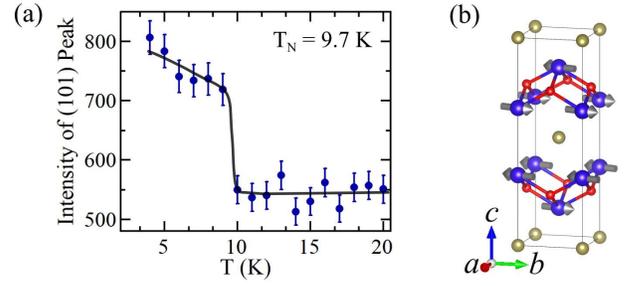}
	\caption {(a) Temperature dependence of
		the peak intensity of (1,0,1) from neutron diffraction on Dy$_{2}$O$_{2}$Te. The solid line is guide to eyes.
		(b) The antiferromagnetic structure of Dy$_{2}$O$_{2}$Te as determined by the Rietveld refinement of powder neutron diffraction data at $T=3.5~$K.} \label{ordpara}
\end{figure}

To determine the antiferromagnetic structure of
Dy$_{2}$O$_{2}$Te below $T_N = 9.7~$K and to check possible magnetic transition around 30~K, powder neutron diffraction measurements were
performed at $T=3.5$, 15, $50~$K and room temperature, respectively. 
The temperature-dependent magnetic scattering at the Bragg peaks (0,0,2), (1,0,1) and (1,1,2) 
are shown in Fig. \ref{NPD} (a-c). The temperature
dependence of the magnetic (1,0,1) peak intensity is shown in Fig. \ref{ordpara} (a). 
The intensity increases abruptly below $10~$K, consistent with the antiferromagnetic transition 
at the N\'{e}el temperature revealed by magnetic susceptibility (Fig. \ref{MT}).
No further magnetic transition around 30~K can be detected in our neutron diffraction study.

The integer Miller index of the magnetic Bragg peaks in \ref{NPD} (a-c) indicates  
that the antiferromagnetic structure of the N\'{e}el state below $T_N$ does not increase the crystalline unit cell
and the magnetic propagation vector \textbf{\textit{k}} = (0,0,0).
For Dy$_{2}$O$_{2}$Te, the space group of the crystal structure is $I 4/mmm $.
For magnetic structures with \textbf{\textit{k}} = (0,0,0),
its little group, which is the same as the crystal point group $4/mmm$, 
has ten irreducible representations (IRs), composed of
eight one-dimensional IRs $ \Gamma_{1} \cdots  \Gamma_{8} $ and
two two-dimensional IRs $ \Gamma_{9} $ and $ \Gamma_{10} $
according to the representation analysis \cite{Bertaut.a05871}.
The spin spaces for Dy$^{3+}$ in $4d$ site can be decomposed as
\begin{equation}
\Gamma = 1\Gamma_{2} + 1\Gamma_{7} + 1\Gamma_{9} + 1\Gamma_{10}.
\end{equation}

Among the four IRs of $\Gamma_{2}$, $\Gamma_{7}$, $\Gamma_{9}$ and $\Gamma_{10}$,
the magnetic structure with $ \Gamma_{9} $ is the best fit with
R$ _{p}$=3.55, R$ _{wp}$=4.36 and $ \chi^{2} $=3.44 for the
neutron powder diffraction pattern measured at 3.5 K [Fig. \ref{NPD} (d)]. 
The Dy neutron absorption correction has been included during the refinement, 
and an impurity phase of 5.6\% in weight is identified in the neutron refinement, consistent with X-ray sample characterization.
The resulting collinear $A$-type antiferromagnetic structure for Dy$_{2}$O$_{2}$Te is depicted in Fig. \ref{ordpara} (b). The ordered moment per Dy ion obtained from Rietveld
refinement is $ \mu $=9.4(1) $ \mu_{B} $/Dy$ ^{3+} $ at 3.5 K. With the tetragonal structure of Dy$_{2}$O$_{2}$Te,
we can only conclude that the moments of the Dy ions are oriented within the $ab$-plane
in this powder diffraction investigation \cite{Shirane.a02507}. On the other hand, the calculations using the point-charge model by McPhase program may provide a clue for the in-plane easy-axis. Our calculation using the $singleion$ module in McPhase indicate the easy-axis of Dy$_2$O$_2$Te could be the $a$-axis or the crystallographically equivalent $b$-axis.

\begin{figure*}[hbtp]
	\centering
	\includegraphics[width=0.9\textwidth]{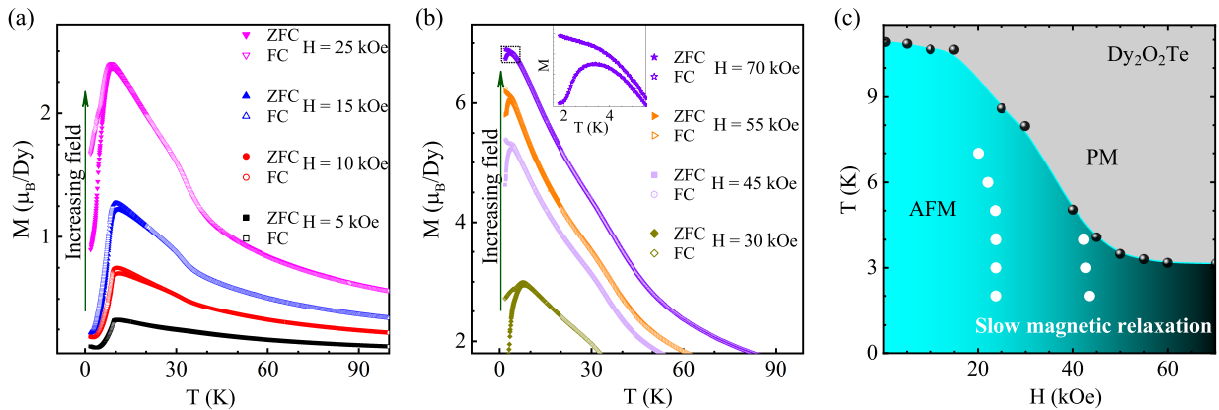}
	\caption {(a) and (b): Magnetic susceptibility versus temperature
		of Dy$_{2}$O$_{2}$Te under different field. The solid and open
		symbols represent the results of ZFC and FC measurements
		respectively. (c) The $H-T$ phase diagram of Dy$_{2}$O$_{2}$Te.
		The black spheres represent the cusp point of ZFC magnetic
		susceptibility. The white points represent the the maxima of dM/dH obtained in Fig. 3.} \label{phasemap}
\end{figure*}

The determination of magnetic structure provides important information in
understanding the novel properties in the {\it R}$_{2}$O$_{2}${\it X} ({\it R}=rare earth, {\it X}=Te, Bi, Sb) family of materials. 
The antiferromagnetic structure of Dy$_{2}$O$_{2}$Te shown in Fig. \ref{ordpara} (b) reveals antiferromagnetic exchange coupling between the 
nearest-neighbor Dy ions and ferromagnetic exchange coupling between the next-nearest-neighbor Dy ions within the Dy$_{2}$O$_{2}$ sublayer. 
The Te layer mediates another antiferromagnetic coupling between the neighboring Dy$_{2}$O$_{2}$ sublayers. 
For superconducting Dy$_{2}$O$_{2}$Bi of the same crystal structure, magnetic transition disappears and superconducting transition emerges 
by excess oxygen incorporation as shown from the resistivity and susceptibility study \cite{Bi221_SC3}. 
Therefore the antiferromagnetic order of Dy$_{2}$O$_{2}$Te is possibly the competing order for the superconducting one in Dy$_{2}$O$_{2}$Bi. 
To verify this speculation, future studies may focus on how the antiferromagnetic order of semiconducting
Dy$_{2}$O$_{2}$Te evolves through Bi-doping to the superconductor, as in the heavy fermion and Fe-based isostructural superconductors.
Further investigations are also needed to check whether spin fluctuations exist in Dy$_{2}$O$_{2}$Bi with inelastic neutron scattering or the NMR techniques.

The material family of {\it R}$_{2}$O$_{2}$${\it X}$ contains many members of antiferromagnet and superconductor. 
In addition to the Dy-series discussed above, we have also synthesized Tb$_{2}$O$_{2}$Te sample and find it undergoes an antiferromagnetic transition 
at $T=2.7~$K. According to previous reports\cite{Tb221,Bi221_SC3}, through Bi substitution for Te, superconductivity at around $2~K$ coexists with antiferromagnetism at $T_N=15~K$ in Tb$_{2}$O$_{2}$Bi. A fertile field is indeed opened for systematical investigations of {\it R}$_{2}$O$_{2}$${\it X}$ compounds.

\subsection{Field-induced slow magnetic relaxation}

\begin{figure}[bthp] \centering
	\includegraphics[width=7.5cm]{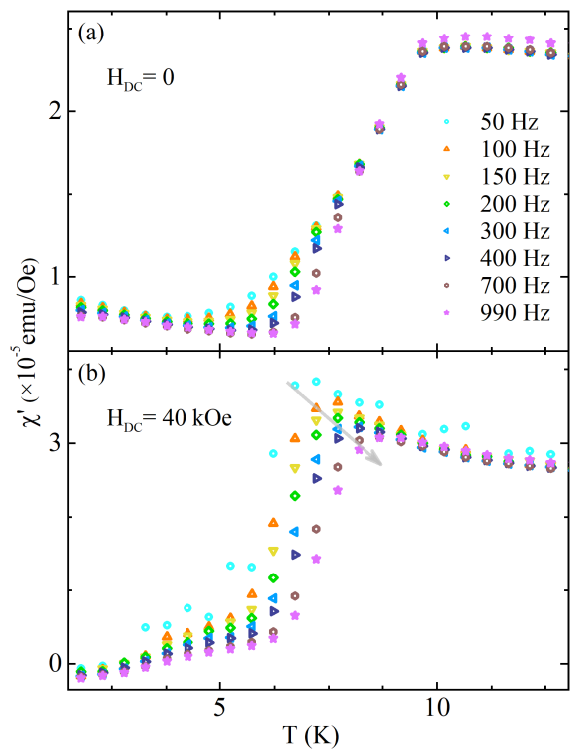}
	\caption {(a) The temperature dependence of the real part of AC susceptibility $\chi'$ under zero DC field and different frequencies. (b) Similar AC susceptibility measurements under DC field of $40~$kOe.} \label{acsus}
\end{figure}

\begin{figure}[bthp] \centering
	\includegraphics[width=7.5cm]{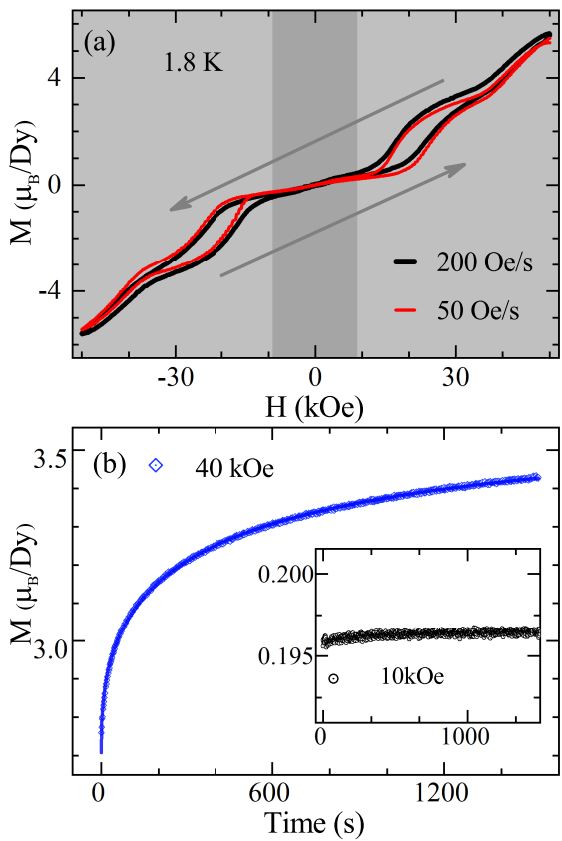}
	\caption { (a) Hysteresis loops of Dy$_{2}$O$_{2}$Te at $T=1.8~$K measured with different field-sweeping-rate. (b) Time dependent of magnetization when the magnetic field is fast settled to $H=40~$kOe at $T=1.8~$K. The inset shows similar measurement for $H=10~$kOe.} \label{MHloop}
\end{figure}

\begin{figure}[bthp] \centering
	\includegraphics[width=7.5cm]{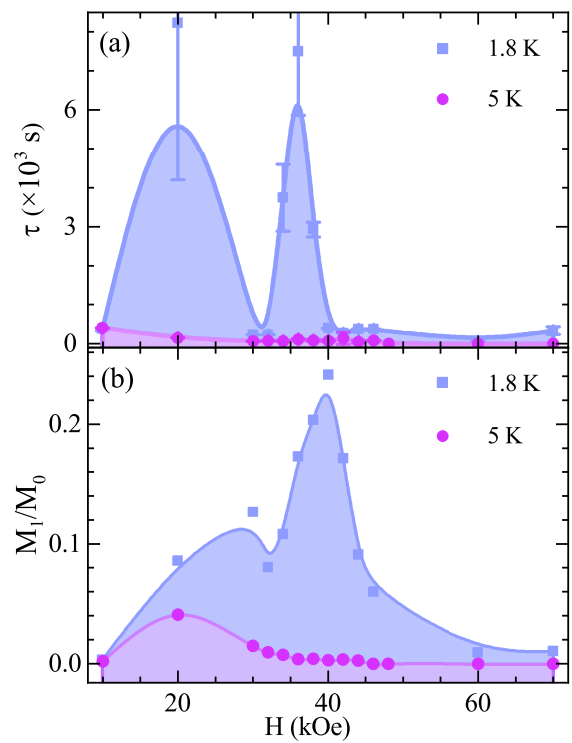}
	\caption {Field and temperature dependent magnetic relaxation time $\tau$ (a) and relaxation ratio $M_1/M_0$ (b). The data were obtained by fitting the time dependent magnetization as described in the text.} \label{magrelax}
\end{figure}

The temperature dependence of magnetization at higher magnetic
fields reveal further anomalous features. As shown in Fig. \ref{phasemap}, the antiferromagnetic-like cusp gradually shifts to lower temperature with increasing field. The field dependent cusp temperatures are plotted in Fig. \ref{phasemap} (c) which marks the border between antiferromagnetic and paramagnetic states. The white points represent the maxima of $dM/dH$ obtained in Fig. \ref{MH} and field-induced canted antiferromagnetic phases may exist in these regions. For $H\leq15~$kOe, the
susceptibility in zero-field-cooling (ZFC) mode almost overlaps
with that in field-cooling (FC) mode. However for $H=25~$kOe, a
separation between ZFC and FC curves occurs below the antiferromagnetic
transition temperature. For $H\geq30~$kOe, the bifurcation between
the ZFC and FC magnetization becomes more obvious (Fig. \ref{phasemap}(b)) and the cusp-feature even disappear in the FC curve. Generally speaking, this behavior strongly resembles the spin frozen in spin glass states\cite{SG1,SG2}. Especially, the AC susceptibility measurement also tracks the peak of real part $\chi'$ shifts to higher temperature with increasing frequency under a DC field of $40~$kOe, while the peak of $\chi'$ does not shift under zero DC field. (illustrated by the arrow in Fig. \ref{acsus} (a) and (b)). However typical spin glass behavior could be easily suppressed by stronger magnetic field\cite{SG1}, the bifurcations between the ZFC and FC susceptibilities, as well as the frequency dependent peak temperature of $\chi'$ are actually induced and enhanced by magnetic field for Dy$_{2}$O$_{2}$Te. Therefore these phenomena may have a different origin.

Fig. \ref{MHloop} (a) presents the hysteresis loop at $T=1.8~$K which has a butterfly-shape. There are big
differences between low- and high-field regions as well. Large
magnetic hysteresis appears at $H\geq12~$kOe while it is absent at
$H\leq12~$kOe. Furthermore, the loops measured under different field-sweeping-rate are not overlap. The above observations strongly suggest the existence of magnetic-field-induced slow magnetic relaxation in Dy$_{2}$O$_{2}$Te. Therefore we directly measured the time dependent magnetization under different field and temperature. As demonstrated in Fig. \ref{MHloop} (b), at $1.8~$K when the magnetic field is fast settled to $H=40~$kOe at $T=1.8~$K, the immediate measurement of sample susceptibility versus time reveals a slow relaxation. This slow
magnetic relaxation behavior is obvious induced by higher magnetic
field as it disappears in a similar measurement under $H=10~$kOe
(inset of Fig. \ref{MHloop} (b)). 

The field-induced slow magnetic behaviors have been widely observed and investigated in single-molecule magnets\cite{SMM1_2003,SMM2_2009,SMM3_2014,SMM4_2020} (many of them are Dy-based complexes) and inorganic compound with diluted Dy ions\cite{1998}. They are explained as a result of magnetic quantum tunneling (MQT) effect. Dy ion has a very large $J$, the Zeeman effect could bring different spin states to the same energy level and generates MQT effect. When the actual field deviates from the tunnelling field, the MQT needs overcome energy barriers and has a low
probability, therefore will simultaneously generates slow magnetic relaxations. However this mechanism mainly applies in material with weak or absent interactions between Dy ions, it seems difficult to work for Dy$_{2}$O$_{2}$Te with long-range magnetic order. Especially when the relaxation time is highly temperature dependent as shown in the following data which contradicts the typical behavior of MQT. So other explanations are needed.

In order to get a quantitative analysis of the slow magnetic behavior, relaxation times $\tau$ were extracted from least squares fits of a
by a stretched exponential equation to the relaxation curves\cite{Slowfit}.

\begin{equation}
	M(t) = M_{0} - M_{1} {e^{-(\frac{t}{\tau})^{\beta}}}.
\end{equation}

In this equation, $\tau$ and $\beta$ are the fitting parameters. Since $M(0)=M_0-M_1$ and $M(\infty)=M_0$, then $M_1/M_0$ can be defined as the ratio of the moments with slow relaxation to the total moments. The fitting results are presented in Fig. \ref{magrelax} (a) and (b). At $1.8~$K, the field-dependent relaxation time $\tau$ have two maximums around $20~$kOe and $38~$kOe. The relaxation ratio $M_1/M_0$ follows roughly the same field-dependent behavior, except the maximums slightly move towards higher field. At $5~$K, both $\tau$ and $M_1/M_0$ are strongly reduced, $M_1/M_0$ has a much weakened maximum at $20~$kOe. If we compare this result with the magnetization isotherms in Fig. 3, one can easily find the close relations between the slow magnetic behaviors and two metamagnetic transitions. At $1.8~$K, the maximum field for $\tau$ and $M_1/M_0$ is only slightly lower than the maximum field at which metamagnetic transitions emerge. With increasing temperature, the slow magnetic relaxations are suppressed similar as that for metamagnetic transitions. 

The two metamagnetic transitions are most likely spin-flop transitions from antiferromagnetic ground state to field-induced canted magnetic states. The above data and analysis strongly suggest the slow magnetic relaxations are actually induced by the transitions from different field-induced magnetic structures. A possible physical origin of this behavior is the phonon bottleneck effect\cite{PB2005,PB2007,PB2018}. This effect is due to the inefficient exchange between the spins and the thermal bath mediated by low-frequency phonons, resulting in slowing of the relaxation process. There have been reports about the phonon bottleneck effect leads to observation of magnetic quantum tunneling, slow magnetic relaxation and butterfly-shaped hysteresis loops\cite{PB2005}, but mainly in molecular complexes. On the other hand, it would also be interesting to check whether this slow magnetic behavior is associated with the possible formation, pinning and movement of new magnetic domains\cite{PCheng_2020}. Overall, this anomalous field-induced slow magnetic behavior has been rarely reported in inorganic correlated compounds. It would be worth for further investigations on its mechanism and whether these properties can be utilized to design novel magnetic devices.

\section{Conclusions}
In summary, the anti-ThCr$_{2}$Si$_{2}$-type compound
Dy$_{2}$O$_{2}$Te is an antiferromagnet with $T_N=9.7~$K. 
A collinear $A$-type antiferromagnetic structure with the magnetic
moment $ \mu $=9.4(1) $ \mu_{B} $/Dy$ ^{3+} $ at 3.5 K confined in the $ab$-plane is identified directly from neutron
diffraction experiments. Further researches are called for on how the family of materials will evolve from antiferromagnetic Dy$_{2}$O$_{2}$Te to superconducting
Dy$_{2}$O$_{2}$Bi. In external magnetic field, anomalous
magnetic properties including bifurcation between ZFC and
FC magnetization, butterfly-shaped magnetic hysteresis and slow magnetic relaxation could be induced in Dy$_{2}$O$_{2}$Te. These properties are
directly associated with the field-induced metamagnetic transitions in Dy$_{2}$O$_{2}$Te. Our results of Dy$_{2}$O$_{2}$Te suggest that the rare-earth {\it R}$_{2}$O$_{2}${\it X} ({\it R}=rare earth, {\it X}=Te, Bi, Sb) compounds is a promising material playground to further explore unconventional superconductivity and antiferromagnetism.

\section*{Acknowledgement}
This work was supported by the National Natural Science Foundation
of China (No. 12074426, No. 11227906, No. 12104255 and No. 12004426), the National Key Research \& Development Projects (No. 2020YFA0406000 and No. 2020YFA0406003), 
and the Fundamental Research Funds for the Central Universities - the Research Funds of Renmin University of China (No. 21XNLG20).

\bibliography{dy221}{}
\end{document}